\tolerance=1000

\centerline{GENERAL-RELATIVISTIC CONSTRAINTS ON THE EQUATION OF STATE
OF DENSE MATTER}
\centerline{IMPLIED BY KILOHERTZ QUASI-PERIODIC OSCILLATIONS IN
NEUTRON-STAR X-RAY BINARIES}
\bigskip

\centerline{W. Klu\'zniak\footnote{$^1$}{Alfred P. Sloan Fellow.}}

Physics Department, University of Wisconsin, Madison, WI 53706, USA

Copernicus Astronomical Center, ul Bartycka 18, 00-716 Warszawa, Poland
\vskip0.5truecm
\centerline{ABSTRACT}
\bigskip\par\noindent
If the observed 
millisecond variability in the X-ray flux of
several neutron-star low-mass X-ray binaries (LMXBs) is interpreted
within a general-relativistic framework 
(Klu\'zniak, Michelson \& Wagoner 1990)
extant at the time of discovery, severe constraints can be placed on
the equation of state (e.o.s.) of matter at supranuclear
densities. The reported maximum frequency ($1.14\pm0.01$ kHz) of
quasiperiodic oscillations observed in sources as diverse as Sco X-1
and 4U 1728-34 would imply that the neutron star masses in these LMXBs
are $M>1.9M_\odot$, and hence many equations of state would be
excluded.  Among the very few still viable equations of state are the
e.o.s.  of Phandaripande and Smith (1975), and e.o.s. AV14 + UVII of
Wiringa, Fiks \& Fabrocini (1988).

\bigskip\par\noindent
{\bf 1. Introduction}
\smallskip\par\noindent
It has long been hoped that observations of neutron stars will allow
an empirical determination of free parameters in the theory of nuclear
interactions as applied to matter at extremely high density.
 Until now, only the softest equations of state of dense matter
had been excluded 
(e.o.s. B, G, in the terminology of Arnett and Bowers [1977])
 by the measured mass of the Hulse-Taylor pulsar
($M=1.44M_\odot$). A third e.o.s., F, has been eliminated  (e.g. 
Cook, Shapiro \& Teukolsky 1994)
 by the probable mass of the binary X-ray pulsar Vela X-1.

 The observed shortest pulsar period, $P=1.56\,$ms, poses no
additional restriction, neither would a 1.2 kHz frequency of stellar rotation,
 because the stiffest e.o.s., M, allows a neutron star period as short as
$P=(1.23\,{\rm kHz})^{-1}=0.83\,$ms, and all other viable equations of
state allow significantly higher rotational frequencies 
(Cook, Shapiro \& Teukolsky 1994).
However, as explained below (Section 4),
identification of an orbital (keplerian) frequency of 1.2 kHz, and 
especially of a {\sl maximum orbital frequency} of 1.2 kHz in the motion
of matter around a neutron star, would
severely constrain hithertofore viable models. It is quite likely that
this frequency has now been observed.

Indeed, it has been proposed  that clumps
in the inner accretion disk of a neutron star which is endowed with at
most a dynamically unimportant magnetic field ($B<10^8$G) will give
rise to millisecond variability in the X-ray flux, and that the
variability will have a characteristic maximum frequency whose
observation should allow a determination of the mass of the neutron
star (Klu\'zniak, Michelson \& Wagoner 1990). Such millisecond
variability has now been observed in the form of kilohertz
quasi-periodic oscillations (kHz QPOs). This letter interprets the
observed kHz QPOs within the general-relativistic framework of
accretion onto neutron stars 
(Lipunov \& Postnov 1984; Klu\'zniak \& Wagoner 1985;
Klu\'zniak \& Wilson 1986; Sunyaev \& Shakura 1986;
Klu\'zniak 1987; Klu\'zniak \& Wilson 1989;
Klu\'zniak, Michelson \& Wagoner 1990; Klu\'zniak \& Wilson 1991; Hanawa 1991;
Klu\'zniak 1991;
Biehle \& Blandford 1993), and describes the resulting limits on the
mass of the neutron star sources of kHz QPOs.

\bigskip\par\noindent
{\bf 2. Neutron stars and accretion disks in LMXBs}
\smallskip\par\noindent
We are going to claim a lower limit to the mass of the compact object in
a class of low-mass X-ray binaries (LMXBs). It is necessary to
review the theory of such sources, to make clear why they are believed to
contain a neutron star and hence why the derived mass-limit is relevant
to the theory of dense matter. The relativistic paradigm of accretion
is also reviewed in this section.

The kHz QPOs were discovered 
in Sco X-1, GX5-1 and several X-ray bursters
(van der Klis {\it et al.} 1996; Strohmeyer {\it et al.} 1996;
Zhang {\it et al.} 1996; Ford {\it et al.} 1997;
van der Klis, M. {\it et al.} 1996.)
These Z, GX, and atoll sources (in current terminology, 
van der Klis 1995) 
are all thought to be binary systems containing an accreting neutron star.
For X-ray bursters, but not for the Z sources,
 there is indeed compelling evidence 
(Lewin \& Joss 1983;
Lewin, van Paradijs, \& Taam 1995)
that they are neutron stars or stars very much like neutron
stars (i.e. with about solar mass, $M\sim M_\odot$, and a stellar
surface at radius $R\sim10^6$cm).
The X-ray bursts 
(Grindlay {\it et al.} 1976; Belian, Conner, \& Evans 1976)
are well explained 
(Maraschi \& Cavaliere 1977; Woosley \& Taam 1976; Joss 1978)
as a result of
 thermonuclear instabilities 
(Hansen \& Van Horn 1975)
in the transmutation of more or
 less freshly accreted material.  Fits to X-ray spectra in the decay
 phase of the burst severely constrain the mass--radius relationship,
 which in all cases is found 
(Lewin,  van Paradijs \& Taam 1993)
to be consistent with theoretical models 
of neutron stars
(Arnett \& Bowers 1977; Friedman, Ipser \& Parker 1986;
Cook, Shapiro \& Teukolsky 1994).

The discovery of kHz QPOs has not been entirely unexpected. There is a
school of thought which would have the magnetic field of at least some
neutron stars in LMXBs so low ($B<10^8$G) as to be of no dynamical
importance.  For certain combinations, strongly dependent
on the equation of state of dense matter, of mass
and angular momentum it would then be possible
for the accretion disk to be terminated
by strong-field effects of general relativity (GR)---this is the
relativistic ``accretion-gap'' regime. It has been thought 
(Klu\'zniak, Michelson \& Wagoner 1990)
that the X-ray flux in such systems is likely to be
modulated at the keplerian frequency in the kHz range.

Specifically, it has been pointed out 
(Klu\'zniak \&  Wagoner 1985),
 that many equations of state of dense matter allow
neutron stars to have radii smaller than the radius, $r_{ms}$,
of the innermost (marginally) stable circular orbit allowed in general
relativity 
(Kaplan 1949).
For such stars 
(Klu\'zniak \&  Wagoner 1985; Klu\'zniak, Michelson \& Wagoner 1990;
Klu\'zniak \& Wilson 1991),
as for black holes 
(Shakura \& Sunyaev 1973; Stoeger 1976; Muchotrzeb \& Paczy\'nski 1982),
the accretion disk will terminate at a radius $r_{in}\sim r_{ms}$.
 (Estimates of the effects of the radiation field, created by emission at the
neutron star surface within $r_{in}$, suggest 
[Biehle \& Blandford 1993]
that
$r_{ms}-r_{in}\approx0.5\,{\rm km}>0$. Under special circumstances,
$r_{ms}<r_{in}$ also seems possible 
[Walker \& M\'esz\'aros 1989].)
Upon crossing the sonic point 
(Stoeger 1976; Muchotrzeb \& Paczy\'nski 1982)
 at $r_s\approx r_{in}$ matter will
fall freely 
(or nearly freely if magnetic or radiative stresses
are not entirely negligible) to 
the surface of the star
(Klu\'zniak \&  Wagoner 1985; Klu\'zniak, Michelson \& Wagoner 1990;
Klu\'zniak \& Wilson 1991).
The free-free cooling time  is shorter than the transit time 
(Klu\'zniak 1987; Klu\'zniak, Michelson \& Wagoner 1990)
so no emission is
expected in the accretion gap between $r_s$ and the top of the
atmosphere in the equatorial accretion belt
(Fig. 1). Some of the phenomenology of the Z
sources can be explained by the ``closing'' of the accretion gap as
the mass accretion rate increases 
(Biehle \& Blandford 1993)

Two clear predictions 
 were made within the accretion-gap framework in GR
(Klu\'zniak, Michelson \& Wagoner 1990; Klu\'zniak \& Wilson 1991).
First, according to self-consistent computations 
of its structure
and radiation field,
the accretion belt emits hard X-rays with a power-law spectrum,
which should be observable at lower mass accretion rates, when the accretion
gap is optically thin to electron scattering (Klu\'zniak \& Wilson 1991).
Similar conclusions were reached by Hanawa 1991.
Second, kilohertz variability
(caused by inhomogeneities in the flow) should
be seen in the X-ray flux with a clearly defined maximum frequency
corresponding to the orbital frequency at the inner edge of the disk 
(Klu\'zniak, Michelson \& Wagoner 1990).
Interestingly, these predictions have now apparently been
verified. In particular, a robust maximum in the QPO
frequency has indeed been observed (Zhang {\it et al.} 
1997).
It seems natural then, to interpret the higher frequency of
the observed kHz QPOs as the keplerian frequency.

\bigskip\par\noindent
{\bf 3. Kilohertz QPOs and orbital frequency}
\smallskip\par\noindent
At present there is only circumstantial evidence as to the nature of the kHz
QPOs. It is therefore worthwile to review the basic
theoretical arguments behind their interpretation.
The idea of QPOs at keplerian frequencies is not new. Twenty five years ago
already, Bath (1973) has argued that ``the periodicities observed in dwarf
novae near maximum light are (...) consistent with periodic eclipses of
temperature fluctuations in the innermost regions of the disk.''
Boyle, Fabian \& Guilbert (1986) argued that QPOs in LMXBs may be caused by
``oscillations of disturbances in the inner regions of a corona above an
accretion disk. Such oscillations will occur predominantly at $\nu_k$, the
keplerian angular frequency of disk matter.'' 
Other models yielding modulation of the flux at keplerian frequency
have also been proposed (e.g. Abramowicz {\it et al.} 1992).

Specifically referring to LMXBs, Klu\'zniak (1987) and
Klu\'zniak, Michelson \& Wagoner (1990) suggested that ``clumps'' in the
inner accretion disk could give rise to QPOs or X-ray modulations at kHz
frequencies. The authors assert that the QPOs
in X-ray pulsars are observed at the keplerian frequency of the inner accretion
disk (this maximum
frequency in a disk which is disrupted by the strong magnetic dipole
is referred to as the ``Ghosh-Lamb'' frequency); hence,
in LMXBs containing only weakly magnetized neutron-stars, the orbital
frequency in the inner disk which terminates close to the marginally stable
orbit of general
relativity should be observable in an identical manner. (In view of
the uncertainties inherent in extrapolating the phenomenology of QPOs by
four orders of magnitude in frequency [from $\sim 0.2\,$Hz in
the X-ray pulsar EXO 2030+375---see Angelini, Stella \& Parmar 1989---to
$\sim1\,$kHz in LMXBs],
the authors also considered the
alternate model of fluctuations in the power spectrum, such as the ones
reported in Cyg X-1).

Thus, the interpretation of QPOs in terms of keplerian frequency was well
established in the literature at the time of discovery of the kHz QPOs.
But, as is well known, two distinct QPO frequencies (each one varying in time)
are observed in many LMXBs, and in the atoll sources the difference of the two
frequencies appears to be constant for any given source
(Strohmeyer {\it et al.} 1996;
Zhang {\it et al.} 1996; Ford {\it et al.} 1997).
Again, the framework for
understanding this phenomenon can easily be found in the literature.
Already Patterson (1979), in discussing the QPO he observed in
AE Aquarii, gives a model which would seem applicable to the kHz QPOs:
``If the blobs [are moving] at their own Keplerian
rotation periods $P_K$ [in] an illuminating searchlight that is
strictly periodic with a period $P_s$ and rotating in the same
direction as the disk, then the result is a QPO with a
period given by
$${1\over P}= {1\over P_s}- {1\over P_K}.$$
Since most of the disk has $P_K>P_s$, we expect in general that QPOs
should have periods slightly longer than $P_s$.''
It is straightforward to apply Patterson's model when two frequencies, with
a constant difference between them, are observed, as in the atoll sources.
 The detection of
the difference frequency, or its harmonic, in X-ray outbursts lends support
to this ``keplerian beat frequency''
model for the lower frequency kHz QPOs, as now all three frequencies
involved are
apparently observed (albeit not all at the same time).

The actual cause of the modulation of flux has yet to be
determined. One possibility is that modulation of flow in the inner
accretion disk is imprinted on the X-ray flux when the flow reaches
the surface of the neutron star. This is the essence of the accretion
gating mechanism proposed to operate at the beat frequency by Alpar \&
Shaham (1985).  The greatest challenge, for any model, is to explain the
timescale of variability of the QPOs. A full theory is unlikely to
emerge before a realistic two-dimensional solution to the accretion
disk in neutron stars is constructed.

In Sco X-1, although there is positive correlation
between the two QPO frequencies, 
their difference varies with time, $\nu_1-\nu_2\ne\,$const, and the third
frequency is not observed
(van der Klis 1996). We may speculate that the second QPO simply corresponds
to an azimuthal velocity in the differentially rotating
equatorial accretion belt.
Calculations show that the infalling stream penetrates the zone of
optical depth unity ($\tau=1$) with a velocity corresponding to
$\nu_0={\rm few}\times10^2$Hz 
(Klu\'zniak \& Wilson 1991).
Inhomogeneities in the flow would rotate about
the star with frequency $\nu_0+\nu_s$, which
would vary as $\nu_0$ varies with differing accretion
conditions.
For reasonable stellar periods, i.e., for $\nu_s^{-1}\equiv P_s\sim\,$few~ms,
$\nu_0+\nu_s$ would correspond to $\sim 0.8\,$kHz.

In summary, without undertaking modeling of the detailed properties of the
QPOs, we may say that it appears likely (Zhang, Stohmyer \& Swank 1997)
that the higher frequency
QPO is the keplerian frequency in the inner accretion disk terminated
by strong-field effects of general relativity, as discussed in
Klu\'zniak, Michelson \& Wagoner (1990). 
The lower frequency QPO appears to be a signature of the presence
of the stellar surface or the magnetic field anchored in it.
On this interpretation, one would only expect a {\sl single} 
QPO in an accreting black hole, e.g., at $\sim 0.3\,$kHz for a $7M_\odot$
Schwarzschild metric. Such a QPO has apparently been
observed 
(Remillard 1997) in the black-hole candidate 1655-40 at the expected
frequency.  This provides additional support for the idea that the QPO
frequency is the orbital frequency.  If this interpretation is
confirmed, the absence or presence of a {\it second} high-frequency
QPO would be the long-awaited direct signature for the absence or
presence of a stellar surface in black-hole or neutron-star candidate
systems, such as 1655-40 and Sco X-1, respectively.

\bigskip\par\noindent
{\bf 4. Constraints on the equation of state}
\smallskip\par\noindent
Let us now follow the prescription of 
Klu\'zniak, Michelson \& Wagoner (1990)
for measuring the mass of the neutron
star through the observation of a maximum frequency
in the power spectrum. This will lead to severe constraints on the
equation of state of dense matter.

The orbital frequency (Fig. 2), seen by an observer
at infinity, has been computed by Klu\'zniak \&  Wagoner 1985
through first order in the dimensionless stellar angular momentum
$j=cJ/GM^2$ 
in the Hartle (1976) metric appropriate to the exterior of a rotating neutron
 star; 
 see also
Klu\'zniak, Michelson \& Wagoner 1990.  
It is
$\Omega(r,M,j)=(GM/r^3)^{1/2}[1-(rc^2/GM)^{-3/2}j].$
Note that fully relativistic models of rotating neutron stars 
(Cook, Shapiro \& Teukolsky 1994)
  allow only $j\le0.6$,
so this first order (in $j$) result should be a good approximation.
The radius of the marginally stable orbit is 
(Klu\'zniak \&  Wagoner 1985)
         $$r_{ms}=6M[1-(2/3)^{3/2}j](G/c^2).
\eqno(1)$$
The frame-dragging term (proportional to $j$) gives at most
a 4\% correction to $\Omega$ at any fixed radius $r>6GM/c^2$,
 however it is not negligible in its effect on the maximum orbital frequency,
which is
$f_{max}=(2\pi)^{-1}\Omega(r_{ms},M,j)=(2\pi)^{-1}
[1+(11/12)(2/3)^{1/2}j](GM/r_{ms}^3)^{1/2}$, i.e. 
(Klu\'zniak, Michelson \& Wagoner 1990),
$$f_{max}=(1+0.749j)({M_\odot/M})\times2.20\,{\rm kHz}.
\eqno(2)$$
The possible range of $j$, for a given rotational frequency of the star,
$1/P$, can be gleaned from Fig. 3, which shows a linear
(in $j$) extrapolation 
(Klu\'zniak 1987)
of the nonrotating models of Arnett \& Bowers
(1977) 
for the canonical mass
 $M=1.4M_\odot$.
The softest hitherto viable e.o.s is A, the stiffest is M. Comparison with
precise numerical calculations of rotating models 
(Cook, Shapiro \& Teukolsky 1994)
shows that the curves labeled A and M
are accurate to about 10\% throughout the range plotted here,
i.e. for $0\le j\le0.5$. Note that a $1.4M_\odot$ neutron star with
rotational frequency of $1/P\approx 330\,$Hz would have
angular momentum roughly in the range $0.1<j<0.3$.
However, the functional relationship $j(P)$ is
quite sensitive to the mass of the star.

Let us now compare the highest frequency of
the kHz QPOs, $\nu_{max}$, with the maximum orbital frequency $f_{max}$ of
eq. (2). In doing so we effectively assume that the disk terminates at
$r_{in}=r_{ms}$. When sufficiently reliable calculations of $r_{in}$ are
carried out, this discussion may have to be
repeated for a different value of $r_{in}$. However, we expect the
conclusions to be upheld because the observed $\nu_{max}$
has a very similar value in Sco X-1 and in the lower luminosity X-ray bursters,
so probably the effective (smallest) value of $r_{in}$ does not
vary much with the accretion rate and hence must be close to $r_{ms}$.

Interpreting, then, the maximum frequency
 of the QPOs, reported to be $\nu_{max}=1.15\,$kHz in each of the
sources\footnote{$^2$}{Using the frequency
reported for 4U 1636-53 (1.17 kHz), or the one reported for Sco~X-1 (1.13 kHz),
we obtain the same constraints on the equation of state.}
4U 1728-34, 4U 1735-44, 4U 0614+091 (Zhang, Strohmyer \& Swank 1997),
as the maximum frequency of stable circular
orbital motion allowed in general relativity, 
$f_{max}$ of eq. (2), we
obtain
$$M=1.91M_\odot\times(1+0.75j).
\eqno (3)$$
This constraint rules out the e.o.s. A, E, D, FPS, UT, as well as
the stiffest e.o.s. M. This is because
the value $1.9M_\odot$ is higher than the
highest mass of a nonrotating star ($j=0$) allowed
 by these hitherto viable equations of state
 and  there are also no rotating
 models (Cook, Shapiro \& Teukolsky 1994), satisfying eq. (3),
constructed with these same equations of state---the only models
(for these e.o.s.) with $M>1.91M_\odot$ require $j\ge0.4$, raising
the mass limit of eq. (3) to $2.5M_\odot$,
which is above the maximum mass limit for any
stable configuration constructed with e.o.s A, E, D, FPS, UT or M.
Thus, regardless of the actual period of the neutron star, only
a few equations of state would be
allowed,
 such as AU (e.o.s. AV14 + UVII,
Wiringa, Fiks \& Fabrocini 1988)
and L (mean field, Phandaripande \& Smith 1975),
which do admit models 
satisfying eq. (3). As it turns out, the allowed e.o.s., L and AU,
admit also models of non-rotating stars with $1.9M_\odot$.
 Note that we would have been unable to constrain
the equation of state in the Schwarzschild metric, the frame-dragging terms
in the Hartle-Thorne metric were crucial to our argument.

\bigskip\par\noindent
{\bf 5. Final remarks}
\smallskip\par\noindent
The discovery of kHz QPOs in many LMXBs
shows that flow close to the neutron stars is not controlled
by magnetic fields in these sources. Standard formulae yield an upper limit
on the stellar magnetic dipole moment of $\mu\le 2\cdot10^{26}$G$\,{\rm cm}^3$
(i.e., a surface dipole field of $\le 2\cdot10^8$G, for a radius of $10^6$cm).
Although both this result, and the magnetic dipole moments of pulsars
inferred from observation of the rotation period and its derivative, are
really order of magnitude estimates, there is at present no basis for asserting
that the observed millisecond pulsar population is similar in the
strength or range of their magnetic field to the LMXB population.
The supposed evolutionary connection between LMXBs and millisecond pulsars
appears to be more tenuous than ever.

In the above remarks, and in Section 4, I have assumed that the higher
QPO frequency is the keplerian frequency. On this assumption, a maximum
frequency is expected (Klu\'zniak, Michelson, Wagoner 1990), as is indeed
observed. At present there is no other framework within which to understand
why a maximum frequency occurs in a continuous distribution of QPO
frequencies, and why its value is so similar in sources as disparate as
Sco X-1 and 4U 1728-34 (Zhang, Strohmyer \& Swank 1997).
On the other hand, there is as yet no direct
evidence that this interpretation is unique---it is as yet too early
to assert that strong-field effects of general relativity have unambigously
been identified. One can hope that future observations and further
theoretical work will lead to absolute certainty.

\bigskip\par\noindent
{\sl Note added in manuscript}
\smallskip\par\noindent

Other authors (Kaaret, Ford \& Chen 1997; and Zhang, Strohmyer \&
Swank 1997) have arrived at the same interpretation of the QPOs and have
derived similar values for the mass of the neutron star.
Aplying the same formalism (our eq. [2]) to the source 4U~1636-53,
in whose spectrum an absorption feature had been reported,
 Kaaret, Ford \& Chen (1997)
have shown how to constrain the equation of state under the additional
two assumptions that the stellar rotation rate and the surface redshift
are known.
Their conclusion that the ``result is consistent with the
mass-radius relation for the AV14~+~UVII equation of state'' agrees with
ours, even though we have used no information at all about the actual spin
rate or the radius of the star and our discussion could have ignored the
source 4U 1636-53 altogether. We can turn the argument around to state
that there is  remarkable agreement between, on the one hand,
 the mass-radius relation of
the few theoretical  models of neutron stars which do allow orbital
frequency in the marginally stable orbit to be as low as
the maximum reported QPO frequency in Sco~X-1,
and, on the other hand,
 the mass-radius relation inferred from the apparent redshift of
FeXXV absorption feature (Waki {\it et al.} 1984, Inoue 1988) in the very
different source 4U 1636-53.

I thank Drs. A. Alpar, M. van der Klis, R.V. Wagoner, 
 and W. Zhang for helpful conversations.
Financial support of the Alfred P. Sloan Foundation is acknowledged.
This work was also supported in part by the Committee for Scientific
Research of the Polish Republic (KBN) through grant number 2-P304-01407.

\bigskip
REFERENCES

Abramowicz, M.A., Lanza, A., Spiegel, E. \& Szuszkiewicz, E. 1992,
 Nature 356, 41.

Alpar, M.A. \& Shaham, J. 1985, Nature 316, 239.

Angelini, L., Stella, L., \& Parmar, A.N. 1989, ApJ 346, 906.

Arnett, W.D., \& Bowers, R.L. 1977, ApJSuppl. 33, 415.

Bath, G.T. 1973, Nature Phys. Sc. 246, 84.

Belian, R.D., Conner, J.P., \& Evans, W.D. 1976, ApJ 206, L135.

Biehle, G.T. \& Blandford, R.D. 1993, ApJ 411, 302.

Boyle, C.B., Fabian, A.C. \& Guilbert, P.W. 1986, Nature 319, 648.

Cook, G.B., Shapiro, S.L., \& Teukolsky, S.A. 1994, ApJ 424, 823.

Ford, E., {\it et al.} 1997, ApJ 475, L123.

Friedman, J., Ipser, J.R., \& Parker, L. 1986, ApJ 304, 115.

Grindlay, J.E., {\it et al.} 1976, ApJ 205, L127.

Hanawa, T. 1991, ApJ, 373, 222.

Hansen, C.J., \& Van Horn, H.M. 1975, ApJ 195, 735.

Inoue, H. 1988, in Physics of Neutron Stars and Black Holes, ed. Y. Tanaka
(Tokyo: Universal Academy Press), 235.

Joss, P.C. 1978, ApJ 225, L123.

Kaaret, Ph., Ford, E.C. \& Chen, K. 1997, ApJLett. 480, L27.

Kaplan, S.A. 1949, Soviet Phys.---JETP Letters 19, 951.

Klu\'zniak, W. 1987, PhD Thesis, Stanford University.

Klu\'zniak, W. 1991, in {\sl IAU colloquium 129\/,} C.~Bertout,
S. Collin-Souffrin, J. P. Lasota \& J. Tran Thanh Van eds.,
(Editions Frontieres: Paris), pp. 327-330.

Klu\'zniak, W., Michelson, P., \& Wagoner, R.V. 1990, ApJ 358, 538.

Klu\'zniak, W., \&  Wagoner, R.V. 1985, ApJ 297, 548.

Klu\'zniak, W., \& Wilson, J. R. 1986, Bulletin of the AAS, 18, 928.

Klu\'zniak, W., \& Wilson, J. R. 1989,
 in {\sl 23rd ESLAB Symposium   on Two Topics in X-ray Astronomy\/,}
 J. Hunt, B. Battrick eds.,   (European Space Agency: Paris), p. 477.

Klu\'zniak, W., \& Wilson, J.R. 1991, ApJ 372, L87.

Lewin, W.H.G, van Paradijs, J., \& Taam, R.E. 1993, Space Sci. Rev. 62, 223.

Lewin, W.H.G, van Paradijs, J., \& Taam, R.E. 1995, in X-ray   binaries,
 ed. W.H.G. Lewin, J. van Paradijs, \& E.P.J. van den Heuvel
 (Cambridge:Cambridge Univ. Press), p. 175.

Lewin, W.H.G., \& Joss, P.C.  1983, in Accretion-driven stellar X-ray sources,
  eds. W.H.G. Lewin, \& E.P.J. van den Heuvel
  (Cambridge:Cambridge Univ. Press), p. 393.

Lipunov, V.M. \& Postnov, K.A. 1984, Astrophys. Space Sci. 106, 103.

Maraschi, L. \& Cavaliere, A. 1977, in Highlights in Astronomy,
 ed. E.A. M\"uller, Vol. 4, Part I, p. 127.

Muchotrzeb, B., \& Paczy\'nski, B. Acta Astron. 32, 1.

Patterson, J. 1979, ApJ 234, 978.

Pandharipande, V.R. \& Smith, R.A. 1975, Phys. Lett. B, 59, 15.

Remillard, R. 1997, in 18th Texas Symposium on Relativistic Astrophysics,
 in the press.

Shakura, N.I., \& Sunyaev, R.A. 1973, AA 24, 337.

Sunyaev, R.A. \& Shakura, N.I. 1986, Pis'ma Astron. Zh. 12, 286
 [Sov. Astron. Lett. 12, 117].

Stoeger, W.R. 1976, AA 53, 267.

Strohmeyer, T.E., {\it et al.} 1996, ApJ 469, L9.

van der Klis, M. 1995, in X-ray binaries, ed. W.H.G. Lewin, J. van  Paradijs,
 \& E.P.J. van den Heuvel (Cambridge:Cambridge Univ. Press),  p. 252.

van der Klis, M. {\it et al.} 1996, ApJ 469, L1.

van der Klis, M. {\it et al.} 1996, IAU Circ. 6424.

Waki, I. et al. 1984, PASJ, 36, 819.

Walker, M.A., \& M\'esz\'aros, P. 1989, ApJ 346, 844.

Wiringa, R.B., Fiks, V. \& Fabrocini, A. 1988, Phys. Rev. C, 38, 1010.

Woosley, S.E. \& Taam, R.E. 1976, Nature, 263, 101.

Zhang, W., Strohmeyer, T.E. \& Swank, J.H. ApJLett. 1997, 482, L167.

Zhang, W., Lapidus, I., White, N.E. \& Titarchuk, L. 1996, ApJ 469, L17.
\vfill\eject
FIGURE CAPTIONS
\bigskip

Fig. 1. A neutron star in the relativistic accretion-gap regime.
Strong-field
effects of gravity terminate the accretion disk close to the innermost
stable circular orbit allowed in general relativity---this innermost
orbit has radius
$r_{ms}\sim 10\,$km (eq. [1]). The gap between the disk and the star
is crossed by the accreting fluid 
in free fall. Kinetic energy of flow is dissipated in the luminous
equatorial accretion belt. [The figure is from Klu\'zniak \& Wilson 1989]
\bigskip

Fig. 2. Orbital frequency in a circular orbit is
plotted as a function of the dimensionless circumferential radius
in the Hartle metric
for two values of angular momentum of the neutron star,
$J=GM^2j/c$, corresponding to a nonrotating star ($j=0$) and to a star
with a period of a few milliseconds ($j=0.3$). Note that 
each curve terminates at a certain maximum frequency, $f_{max}$
(eq. [2]), corresponding to the smallest radius, $r_{ms}$, of a stable
circular orbit allowed in general relativity.
[The figure is from Klu\'zniak, Michelson, \& Wagoner 1990]
\bigskip

Fig. 3. The rotational frequencies computed to first order
in the dimensionless angular momentum, $j$,
of neutron stars with mass $1.3M_\odot$ are shown
for four equations of state (Klu\'zniak 1987).
The extreme curves (e.o.s A and e.o.s M)
are accurate to about 10\% and span the whole
range of hitherto viable equations of state.

\bye